\documentclass[12pt]{article}
\usepackage{authblk}
\usepackage{here}
\usepackage{amsmath}
\usepackage{amsfonts}
\usepackage{bm}
\usepackage{graphicx}          
\usepackage{ascmac}
\usepackage{color}
\usepackage{braket}
\usepackage{here}
\usepackage{cite}
\usepackage[top=25truemm,bottom=30truemm,left=20truemm,right=20truemm]{geometry}
\usepackage{afterpage}
\usepackage{cases}
\usepackage{fancybox}
\usepackage{hyperref}
\hypersetup{colorlinks=true, citecolor=blue, linkcolor=blue, linkbordercolor=0 0 1}
\makeatletter
\renewcommand{\theequation}{\thesection.\arabic{equation}}
\@addtoreset{equation}{section}
\makeatother
\newlength{\myh}
\baselineskip=\normalbaselineskip
\newcommand{\cA}{{\mathcal A}}
\newcommand{\cB}{{\mathcal B}}
\newcommand{\cC}{{\mathcal C}}
\newcommand{\cD}{{\mathcal D}}
\newcommand{\cF}{{\mathcal F}}
\newcommand{\cI}{{\mathcal I}}
\newcommand{\cK}{{\mathcal K}}
\newcommand{\cO}{{\mathcal O}}

\newcommand{\bR}{{\mathbb R}}

\newcommand{\bZ}{{\mathbb Z}}
\newcommand{\bGam}{{\boldsymbol \Gamma}}

\newcommand{\cN}{{\mathcal N}}

\newcommand{\vk}{\vec{k}}
\newcommand{\vx}{\vec{x}}
\newcommand{\vy}{\vec{y}}

\newcommand{\wt}{\widetilde}

\newcommand{\nn}{\nonumber}

\def\matt[#1,#2,#3,#4]{\left(%
\begin{array}{cc} #1 & #2 \\ #3 & #4 \end{array} \right)} 

\begin{document}
\title{{\LARGE 
Comments on contact terms and conformal manifolds
in the AdS/CFT correspondence
}}
\author[1,2]{Tadakatsu Sakai}
\author[1]{Masashi Zenkai}
\affil[1]{Department of Physics, Nagoya University, Nagoya 464-8602, Japan}
\affil[2]{Kobayashi-Maskawa Institute for the Origin of Particles 
and the Universe, Nagoya University, Nagoya 464-8602, Japan}

\date{\today}
\maketitle
%
%
\thispagestyle{empty}
\setcounter{page}{0}

\begin{abstract}

We study the contact terms that appear in the correlation functions
of exactly marginal operators using the AdS/CFT correspondence.
It is known that 
CFT with an exactly marginal deformation requires 
the existence of the contact terms 
with
their coefficients having a geometrical interpretation in the context of 
conformal manifolds.
We show that the AdS/CFT correspondence captures properly
the mathematical 
structure of the correlation functions that is expected from the CFT
analysis.
{}For this purpose, we employ holographic RG
to formulate a most general setup in the
bulk for describing an exactly marginal deformation.
The resultant bulk equations of motion are nonlinear and solved perturbatively
to obtain the on-shell action.
We compute three- and four-point functions of the exactly marginal
operators using the GKP-Witten prescription, and show that
they match with the expected results precisely.
The cut-off surface prescription in the bulk 
serves as a regularization scheme for
conformal perturbation theory in the boundary CFT.
As an application, we examine a double OPE limit of the 
four-point functions. 
The anomalous dimensions of double trace operators
are written in terms of the geometrical data
of a conformal manifold.

\end{abstract}

\setcounter{section}{+0}
\setcounter{subsection}{+0}

\newpage

%
%
\section{Introduction}

Marginal operators of CFT are defined as a scalar operator that has
the conformal dimension $\Delta=d$ at a conformal fixed point
with $d$ being the spacetime dimension.
Perturbing the CFT with marginal operator may spoil conformal symmetry.
In fact, the renormalization group(RG) $\beta$-function
for a coupling constant $\lambda^I$ can be computed using
conformal perturbation theory as
\begin{align}
 \beta^I(\lambda)=\frac{d}{d\log\Lambda}\lambda^I
=(\Delta-d)\lambda^I+c_{IJK}\lambda^J\lambda^K+\cO(\lambda^3) \ .
\label{rg:beta}
\end{align}
Here, $\Delta$ is the conformal dimension of a scalar operator 
$\cO_I$ that is added to the CFT as a perturbation. $c_{IJK}$
is the OPE coefficients that are proportional to
the three-point function of the scalar operators.
The $\beta$-function (\ref{rg:beta}) shows that
the marginal couplings with $\Delta=d$ break conformal
symmetry if $c_{IJK}\ne 0$.
The exactly marginal couplings are defined as the subset
of marginal couplings with the vanishing $\beta$-functions
at all order in $\lambda$.
CFT with an exactly marginal deformation
is then charactorized by a fixed surface rather than a fixed point
in the parameter space. The fixed surface is referred to as a conformal
manifold. The conformal manifold admits
a natural metric that is defined by the 
two-point function of the exactly marginal operators \cite{Zam}
\begin{align}
 g_{IJ}(\lambda)=\langle \cO_I(\infty)\cO_J(0)\rangle_\lambda \ .\nn
\end{align}
Here, the suffix $\lambda$ implies that the two-point function
is evaluated with the perturbed CFT action.

It is discussed in \cite{Sei} that
the three-point function of the exactly marginal couplings has a 
nontrivial contact term
\begin{align}
  \langle \cO_I(x)\cO_J(y)\cO_K(z)\rangle_\lambda
=\frac{1}{(x-z)^{2d}}\,\bGam_{K,IJ}(\lambda)\,\delta^d(x-y)
+(\mbox{cyclic permutations}) \ ,
\label{OOO}
\end{align}
with $\bGam_{K,IJ}$ being the Christoffel symbol defined from
$g_{IJ}$. One way for understanding it is to set 
$\lambda^I=\lambda_0^I+\delta\lambda^I$ with $\delta\lambda^I$
regarded as a perturbation about a reference coupling $\lambda_0^I$.
Then, $g_{IJ}(\lambda)=g_{IJ}(\lambda_0+\delta\lambda)$ is 
required to be computed by the conformal perturbation in $\delta\lambda^I$.
More precisely,
\begin{align}
 g_{IJ}(\lambda_0+\delta\lambda)
=\left\langle \cO_I(1)\cO_J(0)\,
\exp\left(-\int d^dx\,\delta\lambda^K\cO_K(x)\right)
\right\rangle_{\lambda_0} \ .
\label{vev0}
\end{align}
The contact terms in the higher-point functions of
the exactly marginal operators are analyzed in detail
in \cite{Kut}.

The purpose of this paper is to revisit these stucutures
using the AdS/CFT correspondence (for a review, see \cite{adscft})
We first formulate a bulk gravity model in $d+1$ dimensions on the 
basis of holographic RG \cite{dVV} (for a review,
see also \cite{FMS}). It is argued that the gravity dual of
the exactly marginal operator is given by a massless scalar
in AdS$_{d+1}$ with a trivial potential term. This guarantees
that the RG $\beta$-function for the marginal couplings 
vanishes.\footnote{
The existence conditions for an exactly marginal
deformation have been discussed extensively so far.
For recent works from the viewpoint of conformal perturbation theory, 
see 
\cite{Gaberdiel:2008fn, Komargodski:2016auf, Bashmakov:2017rko, Behan:2017mwi, Sen:2017gfr}. 
}
This type of models has been studied extensively so far. A typical 
example is given in 
\cite{Liu:1998ty,DHoker:1999kzh}, which
discuss a dilaton-axion system in AdS$_5$ that is dual
to the gauge coupling and the $\theta$-angle in $\cN=4$ super YM 
in four dimensions. 
{}For a review, see \cite{DHoker:2002nbb}.
{}For an analysis of exactly margical deformation within the context of
the AdS/CFT correspondence,
see also \cite{Tachikawa:2005tq, Louis:2015dca, Ashmore:2016oug,Lust:2017aqj}
and \cite{Bashmakov:2017rko}.

Starting with this model, we work out the on-shell action
by solving the equations of motion of the bulk scalars.
These are given by a non-linear equation and can be solved 
perturbatively in the gravitational coupling. Then, the on-shell
action is evaluated in the form of a power series of the boundary
values that specify the Dirichlet boundary condition of the
bulk scalars at a cut-off surface.
It is emphasized that this procedure is interpreted as a conformal
perturbation theory by identifying the boundary value 
with perturbations to a marginal coupling constant.
{}For an early work on conformal perturbation theory in the
context of the AdS/CFT correspondence, see \cite{Berenstein:2014cia}.
Conformal perturbations become singular
when an integrated operator collides with another.
It is standard to regularize these divergences by limiting
the integral region to avoid the collision of
the operators.
It is found that the GKP-Witten prescription \cite{GKP,W}, 
where a cut-off surface is set near the AdS boundary,
plays a role of a regularization scheme for
the holographic conformal perturbation theory.
{}From the on-shell action, we derive the three- and four-point functions 
of the exactly marginal
operators. It is seen that they reproduce exactly the contact terms that
are expected from the CFT analysis.
We also discuss the four-point function in a double OPE limit.
We read off the anomalous dimensions
of double trace operators composed of the exactly marginal
operators as a function of the geometric data
of the conformal manifold.

In this paper, we fix the bulk metric to be an AdS$_{d+1}$ 
background metric with $d$ even, and
treat only the bulk scalars as a 
dynamical field. This implies that the exchange diagrams
due to the stress tensor are missed from the results in the
four-point function. This simplification is valid for most part
of the computations made in this paper, except for an incomplete
analysis of the anomalous dimensions of the double trace
operators. We leave it as a future work to incorporate the effect
from the dynamical, bulk graviton.

This paper is organized as follows. In section 2, we formulate the bulk
model of an exactly marginal deformation on the basis of
holographic RG. Section 3 is devoted to computing
the correlation functions of exactly marginal operators 
by analyzing this model. The appendix summarizes some
useful formulae for bulk-to-bulk and bulk-to-boundary propagators
of a scalar field in AdS$_{d+1}$.

%
\section{Setup}

One of the most efficient ways for describing RG flows in the
context of the AdS/CFT correspondence is to utilize
the Hamilton-Jacobi(HJ) formulation of a bulk gravity theory 
\cite{dVV}. This formalism is demonstrated in more detail
in \cite{FMS}.

We start with a bulk action in $M_{d+1}$, a $(d+1)$-dimensional
bulk spacetime
\begin{align}
\mathbf S
\left[\hat\gamma_{\mu\nu}(x,\tau),\hat\phi^I(x,\tau)\right]
=\int_{M_{d+1}}d^{d+1}X\sqrt{\hat\gamma}\,\Big\{V(\hat\phi)-\hat R_{(d+1)}
+\frac12L_{IJ}(\hat\phi)\hat\gamma^{\mu\nu}\hat\partial_{\mu}\,\hat\phi^I
\hat\partial_{\nu}\,\hat\phi^J
\Big\} \ .
\end{align}
Here $\hat\gamma_{\mu\nu}$ denotes the bulk metric in $M_{d+1}$,
and is regarded as a dynamical field only in this section.
Using ADM decomposition, it becomes
\begin{align}
 ds^2=\hat\gamma_{\mu\nu}\,dX^{\mu}dX^{\nu}
=\hat N^2(x,\tau)d\tau^2+\hat h_{ij}
(x,\tau)(dx^i+\hat\lambda^i(x,\tau)d\tau)(dx^j+\hat\lambda^j(x,\tau)
d\tau) \ ,
\end{align}
with $\tau\ge \tau_0$.
$\hat h_{ij}$ is an induced metric on a $d$-dimensional hypersurface
at a fixed $\tau$.
The hatted quantities mean off-shell without the equations
of motion imposed. 
$M_{d+1}$ has a $d$-dimensional boundary $\Sigma_{d}$ at $\tau=\tau_0$.
We omit writing the boundary terms 
to be added to the action (\ref{Sbulk}) for simplicity.

We solve the bulk equations of motion under the boundary condition
\begin{align}
\hat h_{ij}(x,\tau=\tau_0)=h_{ij}(x) \ ,~~
\hat\phi^I(x,\tau=\tau_0)=\phi^I(x) \ .\nn
\end{align}
By inserting these classical solutions into the action, 
we obtain the on-shell action $S$, which is the functional
of the boundary values. 
In the dictionary of the AdS/CFT correspondence, $S$ is identified
with the generating functional of CFT$_d$.
Use of the HJ formalism in the bulk gravity allows
one to derive an RG equation for the generating functional.
To see this, we note that 
$S$ obeys the HJ equation by construction, which
follows from the Hamiltonian constraint. 
We divide $S$ into the local and non-local parts
\begin{align}
\frac1{2\kappa_{d+1}^2}S[h(x),\phi(x)]
=\frac1{2\kappa_{d+1}^2}S_\mathrm{loc}
[h(x),\phi(x)]-\Gamma[h(x),\phi(x)]\ ,
\end{align}
with
\begin{align}
 S_{\rm loc}[h(x),\phi(x)]=\int d^dx \sqrt{\gamma}\Big[
W(\phi)+\mbox{(derivative terms)}\Big]
\end{align}
Inserting this into the HJ equation and employing a derivative
expansion, we find that $W(\phi)$ is related with the scalar
potential $V(\phi)$ as
\begin{equation}
V(\phi)=-\frac d{4(d-1)}W^2(\phi)+\frac12L^{IJ}(\phi)\partial_IW(\phi)
		\partial_JW(\phi) \ .
\label{VW}
\end{equation}
Here, $\partial_I=\partial/\partial\phi^I$ and
$L^{IJ}=L^{-1}_{IJ}$.
{}Furthermore, it is found that the non-local part $\Gamma$ obeys
the local RG equation \cite{O1991} with the RG $\beta$-function
for the coupling function $\phi^I(x)$ given by
\begin{equation}
\beta^I(\phi)=-\frac{2(d-1)}{W(\phi)}L^{IJ}(\phi)\partial_JW(\phi) \ .
\end{equation}
This shows that the $\beta$-function vanishes iff $W(\phi)$ is
independent of $\phi^I$. It then follows from (\ref{VW}) that
the scalar potential is a constant, being equal to the bulk
cosmological constant of AdS$_{d+1}$.

To summarize, the bulk setup for studying the exactly marginal
deformation of CFT$_d$ is given by the massless scalar
fields propagating in AdS$_{d+1}$ with a trivial scalar potential
\begin{align}
S_{\rm bulk}=\frac{1}{4\kappa_{d+1}^2}
\int d^{d+1}X\,\sqrt{\gamma}\,
G_{IJ}(\phi)\gamma^{\mu\nu}
\partial_\mu\phi^I\partial_\nu\phi^J \ .
\label{Sbulk}
\end{align}
Hereafter, the unhatted field $\phi^I$ denotes an 
off-shell quantity.
The cosmological constant term is neglected
because this is irrelevant to the rest of the analysis.
This model is studied recently in 
\cite{Bashmakov:2017rko} to argue that it provides us 
with the holographic description
of an exactly marginal deformation.
The bulk scalar fields have self-couplings due to $G_{IJ}(\phi)$,
and also are coupled with the bulk metric $\gamma_{\mu\nu}$.
As mentioned before, this is taken to be a non-dynamical,
AdS$_{d+1}$ background metric
\begin{align}
 \gamma_{\mu\nu}dX^\mu dX^\nu=\frac{dz^2+d\vx^{\, 2}}{z^2} \ ,\nn
\end{align}
with $\vx\in\bR^d$. 


\section{Evaluation of on-shell action}

The equation of motion of the bulk scalar field $\phi^I$ reads
\begin{align}
 \Box\phi^I+\Gamma^I_{IJ}(\phi)\,\partial_\mu\phi^J\partial^\mu\phi^K
=0
\ .\nn
\end{align}
Here, $\Box$ is the Laplacian in AdS$_{d+1}$, and $\Gamma^I_{JK}(\phi)$
is the Christoffel symbol for $G_{IJ}(\phi)$.
It is impossible to find the exact solutions of this. 
In this paper, we attempt to
solve it perturbatively in $\alpha\equiv(2\kappa_{d+1}^2)^{1/2}$.
Expand $\phi^I$ around a constant $\varphi^I$ as
\begin{align}
 \phi^I(X)=\varphi^I+\sum_{n\ge 1}\alpha^n\,
\eta_{(n)}^I(X) \ .\nn
\end{align}
Then, the EOM decomposes as 
\begin{align}
 \Box\eta_{(1)}^I&=0 \ ,
\label{eom1}
\\
 \Box\eta_{(2)}^I&=-\Gamma^I_{JK}\,\partial_\mu\eta_{(1)}^J\,
\partial^\mu\eta_{(1)}^K \ ,
\label{eom2}
\\
\Box\eta_{(3)}^I&=-2\Gamma^I_{JK}\,
\partial_\mu\eta_{(2)}^J\,
\partial^\mu\eta_{(1)}^K\,
-\partial_L\Gamma^I_{JK}\,\eta_{(1)}^L\,
\partial_\mu\eta_{(1)}^J\,
\partial^\mu\eta_{(1)}^K\,
\ ,\nn
\\
&~~\vdots \nn
\end{align}
We set $G_{IJ}=G_{IJ}(\phi=\varphi)$ and
$\Gamma^I_{JK}=\Gamma^I_{JK}(\phi=\varphi)$ from now on.
These equations are solved by imposing the boundary conditions
at the cut-off surface $z=\epsilon>0$
\begin{align}
 \eta_{(1)}^I(z=\epsilon,\vx)=J^I(\vx) \ ,~~~
 \eta_{(n)}^I(z=\epsilon,\vx)=0 \ .~(n\ge 2) 
\label{bc}
\end{align}
The boundary value $J^I(\vx)$ is identified with an external
field coupled with the CFT operator that is dual to $\phi^I$.
When interpreting our results along the line of conformal
perturbation theory,
$\varphi^I$ is identified with the reference coupling $\lambda_0^I$
and $J^I$ with a perturbation about it.

(\ref{eom1}) is solved by using the modified bulk-to-boundary
operator $K_\Delta^\epsilon$ as
\begin{align}
 \eta_{(1)}^I(X)=\int d^d y\,K_\Delta^\epsilon(X,\vy)\,J^I(\vy) \ ,
\end{align}
with $\Delta=d$.
The modified propagators are defined to obey an appropriate
boundary condition at the cut-off surface $z=\epsilon$ rather
than $z=0$. A full account of them is given in \cite{Muck:1998rr}.
See also the appendix \ref{mGK} for a review.
(\ref{eom2}) can be solved by using the modified bulk-to-bulk
propagator as
\begin{align}
 \eta_{(2)}^I(X)=\Gamma^I_{JK}\int d^{d+1}Y\sqrt{\gamma(Y)}\,
G_\Delta^\epsilon(X,Y)\,
\partial_\mu\eta_{(1)}^J\,\partial^\mu\eta_{(1)}^K \ .
\label{eta2}
\end{align} 
Note that this obeys the boundary condition (\ref{bc}) because
$G_\Delta^\epsilon(X,Y)$ vanishes at the cut-off surface
by definition.
The higher-order terms $\eta_{(n)}^I(X)$ with $n\ge 3$ can be computed
recursively. By inserting the solutions into $S_{\rm bulk}$, we obtain
the on-shell action $I$ as a functional of $J^I(\vx)$ order-by-order 
in $\alpha$
\begin{align}
 I[J]=\sum_{n\ge 2}\alpha^{n-2}I_n[J] \ .
\end{align}
$I_n[J]$ is given by a functional of $J$ of power $n$.

As an exercise, $I_2$ is evaluated as
\begin{align}
 I_2&=-\frac{1}{2}\,
G_{IJ}\int_{z=\epsilon}d^dx\,\frac{1}{\epsilon^{d-1}}
\,\eta_{(1)}^I\,\partial_z\eta_{(1)}^J
=-\frac{1}{2\epsilon^{d-1}}\,G_{IJ}\int d^dx\, d^dy\, J^I(\vx)J^J(\vy)\,
\cK^\epsilon(|\vx-\vy|) \ .
\end{align}
Here,
\begin{align}
 \cK^\epsilon(|\vx-\vy|)\equiv
\partial_zK_{\Delta=d}^\epsilon(X,\vy)\big|_{z=\epsilon}
=\int \frac{d^dk}{(2\pi)^d}\,e^{i\vk\cdot(\vx-\vy)}\,
k\,\frac{\partial}{\partial z}\log
\left[
z^{d/2}K_{d/2}(z)\right]\Big|_{z=k\epsilon} \ .
\nn
\end{align}
{}For the purpose of obtaining the two-point function,
we need to expand the integrand in $\epsilon$ and extract
a log divergent term in $\epsilon$, which is a non-analytic
function of $k^2$. 
A standard power expansion of the Bessel
function $K_\nu(z)$ in $z$ is ill-defined for $\nu=d/2\in\bZ_{\ge 1}$.
This problem is resolved by defining $K_{d/2}(z)$ as 
$\lim_{\nu\to d/2}K_\nu(z)$.
It is verified that
\begin{align}
 K_{d/2}(z)=\frac{1}{2}\,\Gamma(d/2)&
\left(\frac{z}{2}\right)^{-d/2}
\left[
\sum_{m=0}^{d/2-1}\frac{(-)^m}{m!}\,
\frac{\Gamma(d/2-m)}{\Gamma(d/2)}
\left(\frac{z}{2}\right)^{2m}
\right.
\nn\\
&\left.
+\sum_{m=0}^\infty
\frac{(-)^{d/2-1}}{m!(d/2-1)!(m+d/2)!}
\left(\frac{z}{2}\right)^{2m+d}
\left\{
2\gamma_E+2\log\left(\frac{z}{2}\right)-H_m-H_{m+d/2}
\right\}
\right] \ .
\label{ex:bessel}
\end{align}
Here, $\gamma_E$ is the Euler gamma constant, and $H_m$ is the harmonic
number. 
Then, we find
\begin{align}
 I_2&=-\frac{1}{2}\,\frac{(-)^{d/2-1}}{2^{d-2}(\Gamma(d/2))^2}\,
G_{IJ}\,
\int d^dx\,d^dy\,J^I(\vx)J^J(\vy)\,
\int \frac{d^dk}{(2\pi)^d}\,e^{i\vk\cdot(\vx-\vy)}\,
k^d\log(k\epsilon) +\cdots \ 
\nn\\
&=-\frac{1}{2}\,\cC_d\,G_{IJ}\,
\int d^dx\,d^dy\,J^I(\vx)J^J(\vy)\,
\frac{1}{(\vx-\vy)^{2d}}+\cdots \ ,
\end{align}
with
\begin{align}
 \cC_d=d\frac{\Gamma(d)}{\pi^{d/2}\Gamma(d/2)} \ .\nn
\end{align}
Here, the power divergent terms in $\epsilon$ are neglected 
because these lead to analytic functions in $k^2$, which are
removed by local counter terms.
It follows that the Zamolodchikov metric is given by
\begin{align}
 g_{IJ}=\cC_d\,G_{IJ} \ .
\end{align}

\subsection{Cubic terms in $J$}
\label{cubic}

The cubic term reads
\begin{align}
 I_3[J]&=\frac{1}{2}\,G_{IJ}\int d^{d+1}X\sqrt{\gamma}
\left(
\partial_\mu\eta_{(1)}^I\,\partial^\mu\eta_{(2)}^J
+
\partial_\mu\eta_{(1)}^J\,\partial^\mu\eta_{(2)}^I
\right)
+\frac{1}{2}\,\partial_KG_{IJ}\int d^{d+1}X\sqrt{\gamma}\,
\eta_{(1)}^K\,\partial_\mu\eta_{(1)}^I\,\partial^\mu\eta_{(1)}^J
\ .\nn
\end{align}
The first term vanishes because of (\ref{eom1}) and the boundary condition
for $\eta_{(2)}^I$ in (\ref{bc}).
The second term can be rewritten as a boundary term at $z=\epsilon$.
To see this, we use integration by parts:
\begin{align}
 I_3=
-\frac{1}{2}\,\partial_KG_{IJ}\int_{z=\epsilon}d^dx\,
\frac{1}{\epsilon^{d-1}}\,\eta_{(1)}^K\,\eta_{(1)}^I\partial_z\eta_{(1)}^J
-\frac{1}{4}\,\partial_KG_{IJ}\int d^{d+1}X\sqrt{\gamma}\,
\partial^\mu\eta_{(1)}^K\left\{
\eta_{(1)}^I\partial_\mu\eta_{(1)}^J
+
\eta_{(1)}^J\partial_\mu\eta_{(1)}^I
\right\} \ .\nn
\end{align}
The integrand in the second term is rewritten as a total derivative
because of (\ref{eom1}).
As a consequence, we obtain
\begin{align}
 I_3&=
-\frac{1}{2}\,\partial_KG_{IJ}\int_{z=\epsilon}d^dx\,
\frac{1}{\epsilon^{d-1}}\,\eta_{(1)}^K\,\eta_{(1)}^I\partial_z\eta_{(1)}^J
+\frac{1}{4}\,\partial_KG_{IJ}\int_{z=\epsilon} d^{d}x\,
\frac{1}{\epsilon^{d-1}}\,\partial_z\eta_{(1)}^K\,\eta_{(1)}^I\,\eta_{(1)}^J
\nn\\
&=-\frac{1}{2\epsilon^{d-1}}\,\Gamma_{K,IJ}\,\int d^dx\,d^dy\,
J^I(\vx)J^J(\vx)J^K(\vy)\,
\cK^\epsilon(|\vx-\vy|)
\ .  \nn
\end{align}
Our focus is on the log divergent term
in $\epsilon$. The computation of $I_3$ proceeds exactly in the same
manner as of $I_2$. We find that
\begin{align}
 I_3=-\frac{1}{2}\,\cC_d\,\Gamma_{K,IJ}\,\int d^dx\,d^dy\,
J^I(\vx)\,J^J(\vx)\,J^K(\vy)\,
\frac{1}{(\vx-\vy)^{2d}}+\cdots \ .
\end{align}
Here, the ultra contact terms that come from power divergent terms
in $\epsilon$ are neglected.

As expected, this result implies that the OPE coefficient among
the CFT operators dual to $\phi^I$ vanishes.
{}Furthermore, we find that the functional derivative
$-\delta^2I_3/\delta J^I(\vx)\delta J^J(\vy)\big|_{J=\delta\lambda/\alpha}$
reproduces 
the first-order term of $\delta\lambda$ in (\ref{vev0}).

\subsection{Quartic terms in $J$}

It is found that
\begin{align}
 I_4[J]&=
\frac{1}{2}\,G_{IJ}\int d^{d+1}X\sqrt{\gamma}
\left(
2\,\partial_\mu\eta^I_{(1)}\,\partial^\mu\eta^J_{(3)}
+
\partial_\mu\eta^I_{(2)}\,\partial^\mu\eta^J_{(2)}
\right)
\nn\\
&\qquad+\frac{1}{2}\,\partial_KG_{IJ}\int d^{d+1}X\sqrt{\gamma}
\left(
\eta^K_{(2)}\,\partial_\mu\eta^I_{(1)}\,\partial^\mu\eta^J_{(1)}
+
2\,\eta^K_{(1)}\,\partial_\mu\eta^I_{(2)}\,\partial^\mu\eta^J_{(1)}
\right)
\nn\\
&\qquad+\frac{1}{4}\,\partial_K\partial_LG_{IJ}\int d^{d+1}X\sqrt{\gamma}
\,
\eta^K_{(1)}\,\eta^L_{(1)}\,
\partial_\mu\eta^I_{(1)}\,\partial^\mu\eta^J_{(1)}
\nn\\
&=-\frac{1}{2}\,G_{IJ}\int d^{d+1}X\sqrt{\gamma}\,
\eta^I_{(2)}\,\Box\eta^J_{(2)}
-\Gamma_{K,IJ}\int d^{d+1}X\sqrt{\gamma}\,
\eta^K_{(2)}\,\partial_\mu\eta^I_{(1)}\,\partial^\mu\eta^J_{(1)}
\nn\\
&\qquad
+\frac{1}{4}\,\partial_K\partial_LG_{IJ}\int d^{d+1}X\sqrt{\gamma}
\,
\eta^K_{(1)}\,\eta^L_{(1)}\,
\partial_\mu\eta^I_{(1)}\,\partial^\mu\eta^J_{(1)}
\ .
\end{align}
Here, (\ref{eom1}) and (\ref{bc}) are used.
Inserting the solution (\ref{eta2}) gives
\begin{align}
 I_4[J]=I_{\rm ex}[J]+I_{\rm cont}[J] \ ,\nn
\end{align}
with
\begin{align}
 I_{\rm ex}&=-\frac{1}{2}\,\Gamma_{M,IJ}\Gamma^M_{KL}
\int d^{d+1}X\,d^{d+1}Y\sqrt{\gamma(X)}\sqrt{\gamma(Y)}\,
G_{\Delta}^\epsilon(X,Y)\,
\partial_\mu\eta^I_{(1)}(X)\,\partial^\mu\eta^J_{(1)}(X)
\cdot\partial_\nu\eta^K_{(1)}(Y)\,\partial^\nu\eta^L_{(1)}(Y)
\ ,
\\
I_{\rm cont}&=
\frac{1}{4}\,\partial_K\partial_LG_{IJ}\int d^{d+1}X\sqrt{\gamma}
\,
\eta^K_{(1)}\,\eta^L_{(1)}\,
\partial_\mu\eta^I_{(1)}\,\partial^\mu\eta^J_{(1)}
\ .
\label{Icont}
\end{align}
$I_{\rm ex}$ and $I_{\rm cont}$ are interpreted as the contributions
from an exchange and contact diagram, respectively.

It is verified that the exchange diagram can be rewritten into the
form of the contact diagram (\ref{Icont})
up to a contact term.
This result was first obtained in \cite{Liu:1998ty} in an analysis
of the dilaton-axion system, although no attention is paid
to the contact terms.
To see this, we first perform integration by parts with respect to
$\partial/\partial X^{\mu}$ twice.
Then,
\begin{align}
 I_{\rm ex}&=
\frac{1}{4}\Gamma_{M,IJ}\Gamma^M_{KL}
\int d^{d+1}X\,d^{d+1}Y\sqrt{\gamma(X)}\sqrt{\gamma(Y)}\,
\partial_\mu^{(X)}G_{\Delta}^\epsilon(X,Y)\cdot
\partial^\mu\left(\eta^I_{(1)}(X)\,\eta^J_{(1)}(X)\right)
\cdot\partial_\nu\eta^K_{(1)}(Y)\,\partial^\nu\eta^L_{(1)}(Y)
\nn\\
&=
-\frac{1}{4}\Gamma_{M,IJ}\Gamma^M_{KL}
\int d^{d+1}X\,d^{d}y\sqrt{\gamma(X)}\,
K_\Delta^\epsilon(X,\vy)\,
\partial_\mu\eta^K_{(1)}(X)\,\partial^\mu\eta^L_{(1)}(X)
\,J^I(\vy)\,J^J(\vy)\,
\nn\\
&\qquad+
\frac{1}{4}\Gamma_{M,IJ}\Gamma^M_{KL}
\int d^{d+1}X\sqrt{\gamma}\,
\partial_\mu\eta^K_{(1)}\,\partial^\mu\eta^L_{(1)}
\,\eta_{(1)}^I\,\eta_{(1)}^J\ .
\label{Iex1}
\end{align}
Here, (\ref{Kdelta}) and
\begin{align}
 \Box_X G_\Delta^\epsilon(X,Y)=-\frac{1}{\sqrt{\gamma}}\,
\delta^{d+1}(X-Y)\ ,
\nn
\end{align}
are used. 
The first term in (\ref{Iex1}) can be further rewritten
using integration by parts again and (\ref{Kdelta_eps}) to show
that this is given by a contact term.
The second term in (\ref{Iex1}) takes exactly the same form
of the contact diagram (\ref{Icont}) up to constant prefactors,
as promised. We obtain
\begin{align}
 I_{\rm ex}&=
\frac{1}{4}\,\Gamma_{M,IJ}\Gamma^M_{KL}
\int d^{d+1}X\sqrt{\gamma}\,
\eta_{(1)}^I\,\eta_{(1)}^J
\,\partial_\mu\eta^K_{(1)}\,\partial^\mu\eta^L_{(1)}
\nn\\
&\qquad+
\frac{1}{4\epsilon^{d-1}}^,\Gamma_{M,IJ}\Gamma^M_{KL}
\int d^dx\,d^dy\,
\cK^\epsilon(|\vx-\vy|)\,
J^I(\vx)J^J(\vx)J^K(\vy)\left(
J^L(\vx)-\frac{1}{2}J^L(\vy)\right) \ .
\label{Iex2}
\end{align}

In order to examine if $I_4$ reproduces the contact terms
that are consistent with exactly marginal deformation, 
we study
\begin{align}
 -\alpha\,\frac{\delta^3 I_4}
{\delta J^I(\vx_1)\delta J^J(\vx_2)\delta J^K(\vx_3)}
\bigg|_{J=\delta\lambda/\alpha} \ .
\end{align}
This should be interpreted as the linear correction
in $\delta\lambda$ to the three-point function 
that results from $I_3$.
We start by analyzing the contributions from  
the contact diagram $I_{\rm cont}$.
It is useful to rewrite it as
\begin{align}
 I_{\rm cont}=\frac{1}{4}\,\partial_I\partial_JG_{KL}
\int d^d y_1\cdots d^d y_4\,
J^I(\vy_1)J^J(\vy_2)J^K(\vy_3)J^L(\vy_4)\,
\cF(\vy_1,\vy_2,\vy_3,\vy_4) \ ,
\end{align}
with
\begin{align}
 \cF(\vy_1,\vy_2,\vy_3,\vy_4)
=\int_\epsilon^\infty\frac{dz}{z^{d-1}}\int d^dx\,
K_\Delta^\epsilon(X,\vy_1)\,K_\Delta^\epsilon(X,\vy_2)\,
\partial_\mu K_\Delta^\epsilon(X,\vy_3)\,
\partial^\mu K_\Delta^\epsilon(X,\vy_4)\ .
\end{align}
It is found that
\begin{align}
 &\frac{\delta^3 I_{\rm cont}}
{\delta J^I(\vx_1)\delta J^J(\vx_2)\delta J^K(\vx_3)}
\bigg|_{J={\rm const.}}
\nn\\
&
=J^L\partial_L
\Big[
\partial_IG_{JK}\,f(\vx_1,\vx_2,\vx_3)
+\partial_JG_{KI}\,f(\vx_2,\vx_3,\vx_1)
+\partial_KG_{IJ}\,f(\vx_3,\vx_1,\vx_2)
\Big] \ .\nn
\end{align}
Here, we define 
\begin{align}
 f(\vx_1,\vx_2,\vx_3)
=\int d^dy\,\cF(\vx_1,\vy,\vx_2,\vx_3)
\ ,\nn
\end{align}
and used the relation
\begin{align}
 \int d^dy\,\cF(\vx_1,\vx_2,\vx_3,\vy)=0\ ,\nn
\end{align}
which follows from (\ref{int:K}).
It is not difficult to see that $f(\vx_1,\vx_2,\vx_3)$ is given by
a contact term of the form
\begin{align}
 f(\vx_1,\vx_2,\vx_3)
=-\frac{1}{2\epsilon^{d-1}}
\Big[
\delta^d(\vx_{12})\,\cK^\epsilon(x_{13})
+\delta^d(\vx_{13})\,\cK^\epsilon(x_{12})
-\delta^d(\vx_{23})\,\cK^\epsilon(x_{12})
\Big] \ .\nn
\end{align}
Here, $\vx_{ij}=\vx_i-\vx_j$.
It follows from these results that
\begin{align}
-\alpha\,&\frac{\delta^3 I_{\rm cont}}
{\delta J^I(\vx_1)\delta J^J(\vx_2)\delta J^K(\vx_3)}
\bigg|_{J={\delta\lambda/\alpha}}
\nn\\
&
=\frac{1}{\epsilon^{d-1}}\,\delta\lambda^L\partial_L
\Big[
\delta^d(\vx_{12})\,\cK^\epsilon(x_{23})\,\Gamma_{K,IJ}
+\delta^d(\vx_{23})\,\cK^\epsilon(x_{31})\,\Gamma_{I,JK}
+\delta^d(\vx_{31})\,\cK^\epsilon(x_{12})\,\Gamma_{J,KI}
\Big] \ .
\end{align}

The functional derivative of $I_{\rm ex}$ is computed 
in the same manner. It can be proved that
\begin{align}
\frac{\delta^3 I_{\rm ex}}
{\delta J^I(\vx_1)\delta J^J(\vx_2)\delta J^K(\vx_3)}
\bigg|_{J={\rm const.}}=0 \ ,\nn
\end{align}
due to cancellation of the functional derivative of
the bulk term in (\ref{Iex2}) with that of the boundary
term in (\ref{Iex2}).
Therefore, we find
\begin{align}
-\alpha\,&\frac{\delta^3 I_{4}}
{\delta J^I(\vx_1)\delta J^J(\vx_2)\delta J^K(\vx_3)}
\bigg|_{J={\delta\lambda/\alpha}}
\nn\\
&
=\frac{1}{\epsilon^{d-1}}\,\delta\lambda^L\partial_L
\Big[
\delta^d(\vx_{12})\,\cK^\epsilon(x_{23})\,\Gamma_{K,IJ}
+\delta^d(\vx_{23})\,\cK^\epsilon(x_{31})\,\Gamma_{I,JK}
+\delta^d(\vx_{31})\,\cK^\epsilon(x_{12})\,\Gamma_{J,KI}
\Big] \ .
\end{align}
As expected, this coincides with the linear correction
to the three-point function of the exactly marginal
operators that comes from $I_3$.

\subsection{Double OPE limit and double trace operators}

In this subsection, we examine  the double OPE limit of
the four-point functions of the 
exactly marginal operators with an emphasis on
the structure of double trace operators.
The results given below are a simple extension
of the papers \cite{DHoker:1999kzh, DHoker:1999mic}.

The leading contribution to the four-point functions in $\alpha$
expansions is given by disconnected diagrams. Including them,
the four-point function reads
\begin{align}
&  \langle 
\cO_I(\vx_1)\,\cO_J(\vx_2)\,\cO_K(\vx_3)\,\cO_L(\vx_4)
\rangle
\nn\\
&=
\frac{g_{IJ}g_{KL}}{x_{12}^{2\Delta}x_{34}^{2\Delta}}
+\frac{g_{IK}g_{JL}}{x_{13}^{2\Delta}x_{24}^{2\Delta}}
+\frac{g_{IL}g_{JK}}{x_{14}^{2\Delta}x_{23}^{2\Delta}}
 -\alpha^2\frac{\delta^4 I_4[J]}
{\delta J^I(\vx_1)\delta J^J(\vx_2)\delta J^K(\vx_3)\delta J^L(\vx_4)} \ .
\label{4ptf}
\end{align}
Here, no contact term in the four-point function comes out
because 
the positions of the operators are taken to be different
from each other.
This implies that
we are allowed to neglect the boundary terms at $z=\epsilon$
and set $\epsilon=0$ in $I_4$ before performing the $z$ integral.
Then, we have
\begin{align}
 I_4&=\frac{1}{4}\,A_{IJKL}\,\int d^{d+1}X\sqrt{\gamma}\,
\eta_{(1)}^I\,\eta_{(1)}^J\,
\partial_\mu\eta_{(1)}^I\,\partial^\mu\eta_{(1)}^I\ ,
\nn\\
&=
\frac{1}{4}\,A_{IJKL}
\left(
\frac{\Gamma(d)}{\pi^{d/2}\,\Gamma(d/2)}
\right)^4\,
\int d^dy_1\cdots d^dy_4\,
J^I(\vy_1)J^J(\vy_2)J^K(\vy_3)J^L(\vy_4)\,\,
\cI(\vy_1,\vy_2,\vy_3,\vy_4)\ ,
\end{align}
with
\begin{align}
 A_{IJKL}=\partial_I\partial_JG_{KL}+\Gamma_{M,IJ}\Gamma^M_{KL} \ ,
\end{align}
and
\begin{align}
 \cI(\vy_1,\vy_2,\vy_3,\vy_4)=\int_0^\infty\frac{dz}{z^{d-1}}
\int d^dx
&
\left(
\frac{z}{z^2+(\vx-\vy_1)^2}
\right)^d\,
\left(
\frac{z}{z^2+(\vx-\vy_2)^2}
\right)^d\,
\nn\\
&\times
\delta^{\mu\nu}\,
\frac{\partial}{\partial X^\mu}
\left(
\frac{z}{z^2+(\vx-\vy_3)^2}
\right)^d\,
\frac{\partial}{\partial X^\nu}
\left(
\frac{z}{z^2+(\vx-\vy_4)^2}
\right)^d\ .
\end{align}
Here, (\ref{K:eps0}) is used.
It then follows that
\begin{align}
& -\frac{\delta^4 I_4[J]}
{\delta J^I(\vx_1)\delta J^J(\vx_2)\delta J^K(\vx_3)\delta J^L(\vx_4)} 
\nn\\
=&-
\left(
\frac{\Gamma(d)}{\pi^{d/2}\,\Gamma(d/2)}
\right)^4\,
\Big(
A_{IKJL}\,\cI(\vx_1,\vx_3,\vx_2,\vx_4)
+A_{JLIK}\,\cI(\vx_2,\vx_4,\vx_1,\vx_3)
\nn\\
&\hspace{3.5cm}
+A_{IJKL}\,\cI(\vx_1,\vx_2,\vx_3,\vx_4)
+A_{KLIJ}\,\cI(\vx_3,\vx_4,\vx_1,\vx_2)
\nn\\
&\hspace{3.5cm}
+A_{ILJK}\,\cI(\vx_1,\vx_4,\vx_2,\vx_3)
+A_{JKIL}\,\cI(\vx_2,\vx_3,\vx_1,\vx_4)
\Big) \ .
\end{align}
It is useful to rewrite $\cI$ by employing the $D$-function 
defined in \cite{DHoker:1999kzh}
\begin{align}
 D_{\Delta_1\Delta_2\Delta_3\Delta_4}
(\vx_1,\vx_2,\vx_3,\vx_4)
=\int_0^\infty\frac{dz}{z^{d+1}}\int d^dx\,
\prod_{i=1}^4
\left(
\frac{z}{z^2+(\vx-\vx_i)^2}
\right)^{\Delta_i}\ ,
\end{align}
Using the formulae summarized in the appendix gives
\begin{align}
 \cI(\vx_1,\vx_2,\vx_3,\vx_4)
&=
d^2\Big(
D_{dddd}(\vx_1,\vx_2,\vx_3,\vx_4)-2x_{34}^2
D_{dd,d+1,d+1}(\vx_1,\vx_2,\vx_3,\vx_4)\Big) 
\nn\\
&=
d^2\left(
1+\frac{3}{d}\,x_{34}^2\,\frac{\partial}{\partial x_{34}^2}
\right) 
D_{dddd}(\vx_1,\vx_2,\vx_3,\vx_4)\ .
\end{align}
The $D$-function
depends on two conformal invariants defined as
\begin{align}
s=\frac{1}{2}\,\frac{x_{13}^2\,x_{24}^2}
{x_{12}^2\,x_{34}^2+x_{14}^2\,x_{23}^2}   \ ,~~~
t=\frac{x_{12}^2\,x_{34}^2-x_{14}^2\,x_{23}^2}
{x_{12}^2\,x_{34}^2+x_{14}^2\,x_{23}^2}   \ .
\nn
\end{align}

A double OPE limit is defined as
\begin{align}
 |x_{13}|\ll |x_{12}| \ ,~~~
 |x_{24}|\ll |x_{12}| \ .\nn
\end{align}
In this limit, the conformal invariants become
\begin{align}
 s\sim \frac{1}{4}\,\frac{x_{13}^2\,x_{24}^2}{x_{12}^4}\to 0 \ ,~~~
t\sim
-\frac{1}{x_{12}^2}\left[
\vx_{13}\cdot\vx_{24}-2\,\frac{(\vx_{12}\cdot \vx_{13})(\vx_{12}\cdot \vx_{24})}
{x_{12}^2}
\right]\to 0\ ,
\nn
\end{align}
with $s/t^2$ kept finite.
As shown in \cite{DHoker:1999kzh},
this limit gives rise to finite and log divergent terms in the $D$-function:
\begin{align}
 D_{dddd}(\vx_1,\vx_2,\vx_3,\vx_4)\sim
-\frac{\pi^{d/2}\,\Gamma(3d/2)}{2\,\Gamma(2d)}\,
\frac{1}{x_{12}^{4d}}\,
\Big(
3H_{d-1}-H_{d-1/2}+\log s
\Big) \ .\nn
\end{align}
Then, the four-point function
in the double OPE limit becomes
\begin{align}
&  \langle 
\cO_I(\vx_1)\,\cO_J(\vx_2)\,\cO_K(\vx_3)\,\cO_L(\vx_4)
\rangle
\sim
\frac{g_{IK}g_{JL}}{x_{13}^{2\Delta}x_{24}^{2\Delta}}
+
\frac{g_{IJ}g_{KL}+g_{IL}g_{JKL}}{(x_{12}^2)^{2\Delta}}
\nn\\
+&\alpha^2\,
\frac{d^2}{2\pi^{3d/2}}\,\frac{\Gamma(3d/2)}{\Gamma(2d)}\,
\left(\frac{\Gamma(d)}{\Gamma(d/2)}\right)^4\,
\frac{1}{(x_{12}^2)^{2d}}
\bigg[
\left(3H_{d}-H_{d-1/2}\right)\cA_{IKJL}
-5\left(3H_{d-1}-H_{d-1/2}\right)\cB_{IKJL}
\nn\\
&
\hspace{7cm}
+\left(\cA_{IKJL}-5\,\cB_{IKJL}\right)\log s
\bigg]\ ,
\label{4pt1}
\end{align}
with
\begin{align}
\cA_{IKJL}=A_{IKJL}+A_{JLIK} \ ,~~~
\cB_{IKJL}=A_{IJKL}+A_{KLIJ}+A_{ILJK}+A_{JKIL} \ .\nn
\end{align}

We interpret this result in terms of the OPE of the exactly
marginal operators
\begin{align}
 \cO_I(x_1)\cO_K(x_3)\sim
\frac{g_{IK}}{x_{13}^{2\Delta}}
+\sum_{a}
\frac{C_{IK}^{~~~a}}{x_{13}^{2\Delta-\Delta_{a}}}\,\cO_{a}(x_1)
\ .
\label{OPE}
\end{align}
Some comments about this are in order.
We neglect the terms in the RHS that depends on the stress 
tensor for simplicity, because these are of no interest
in this paper.
As shown in section \ref{cubic}, the OPE coefficients
among the exactly marginal operators vanish.
$\{\cO_{a}\}$ denotes the whole set of  
the double trace operators that are defined by the product
of the exactly marginal operators.
In this paper, we choose the basis of the double trace operators 
in such a way that the two-point functions are orthonormal:
\begin{align}
 \langle \cO_{a}(x_1)\cO_{b}(x_2)\rangle
&=
\frac{\delta_{ab}}
{x_{12}^{2\Delta_{a}}} \ .\nn
\end{align}
$\Delta_{a}$ is the conformal dimensions of $\cO_{a}$,
and related
to $\Delta$ as
\begin{align}
 \Delta_{a}=2\Delta+\gamma_{a} \ ,
\nn
\end{align}
with $\gamma_{a}$ equal to the anomalous dimension 
of $\cO_{a}$.
$C_{IK}^{~~~a}$ is the OPE 
coefficient among the
exactly marginal operators and the double trace operators.
Using the OPE (\ref{OPE}), the four-point function becomes
\begin{align}
\langle 
\cO_I(\vx_1)\,\cO_J(\vx_2)\,\cO_K(\vx_3)\,\cO_L(\vx_4)
\rangle
\sim
\frac{g_{IK}g_{JL}}{x_{13}^{2\Delta}x_{24}^{2\Delta}}
+
\frac{1}{(x_{12}^2)^{2\Delta}}
\sum_{a}
C_{IK}^{~~~a}C_{JL}^{~~~a}\,
(4s)^{\gamma_{a}/2} 
\ .
\label{4pt2}
\end{align}

Now we equate (\ref{4pt1}) with (\ref{4pt2}) to relate
the OPE coefficients and the anomalous dimensions
of the double trace operators with the bulk data
given in (\ref{4pt1}).
This is possible if the conformal dimensions of the single
trace operators $\cO_I$ and $T_{ij}$ receive no
$\alpha$ correction. The stress tensor is not renormalized
because it is conserved.
We assume that $\cO_I$ remains to be an exactly marginal operator
beyond the planar limit.
See \cite{Bashmakov:2017rko}
for a discussion about it
from the holographic
viewpoint. 
We furthermore assume that 
$\gamma_{a}=\cO(\alpha^2)$
and the OPE coefficients $C_{IJ}^{~~~a}$ receive an $\cO(\alpha^2)$
correction
\begin{align}
 C_{IJ}^{~~~a}=C_{IJ}^{(0)\,a}+\alpha^2\,C_{IJ}^{{(1)}\,a} \ .
\nn
\end{align}
{}For instance, the bulk AdS$_5$ has $\alpha\propto 1/N$ so that 
the anomalous dimensions are of $\cO(1/N^2)$.
Then, by expanding the R.H.S. of (\ref{4pt2}) in $\alpha$, 
we find
\begin{align}
&\langle 
\cO_I(\vx_1)\,\cO_J(\vx_2)\,\cO_K(\vx_3)\,\cO_L(\vx_4)
\rangle
\sim
\frac{g_{IK}g_{JL}}{x_{13}^{2\Delta}x_{24}^{2\Delta}}
\nn\\
&+
\frac{1}{(x_{12}^2)^{2\Delta}}
\sum_{a}\left[
C_{IK}^{(0)\,a}C_{JL}^{(0)\,a}
+\alpha^2
\left(
C_{IK}^{(0)\,a}C_{JL}^{(1)\,a}
+C_{IK}^{(1)\,a}C_{JL}^{(0)\,a}
\right)
+C_{IK}^{(0)\,a}\,\gamma_{a}\,C_{JL}^{(0)\,a}\,\log 2
\right]
\nn\\
&+
\frac{1}{2}\,\frac{\log s}{(x_{12}^2)^{2\Delta}}
\sum_{a}
C_{IK}^{(0)\,a}\,\gamma_{a}\,C_{JL}^{(0)\,a}
+\cO(\alpha^4)\ .
\label{4pt3}
\end{align}
Comparing (\ref{4pt1}) with (\ref{4pt3}) yields
\begin{align}
\sum_{a}C_{IK}^{(0)\,a}C_{JL}^{(0)\,a}&=
g_{IJ}g_{KL}+g_{IL}g_{JK} \ ,
\label{CC1}
\\
\sum_{a}C_{IK}^{(0)\,a}\,\gamma_{a}\,C_{JL}^{(0)\,a}&=
2\alpha^2\,\frac{d^2}{2\pi^{3d/2}}\,
\frac{\Gamma(3d/2)}{\Gamma(2d)}
\left(\frac{\Gamma(d)}{\Gamma(d/2)}\right)^4\,
\left(\cA_{IKJL}-5\,\cB_{IKJL}\right)
\ ,
\label{CC2}
\\
\sum_a\left(
C_{IK}^{(0)\,a}C_{JL}^{(1)\,a}
+C_{IK}^{(1)\,a}C_{JL}^{(0)\,a}
\right)
&=
\frac{d^2}{2\pi^{3d/2}}\,
\frac{\Gamma(3d/2)}{\Gamma(2d)}
\left(\frac{\Gamma(d)}{\Gamma(d/2)}\right)^4
\nn\\
&\hspace{-2cm}\times
\Big[
\left(3H_{d}-H_{d-1/2}-\log 4\right)\cA_{IKJL}
-5\left(3H_{d-1}-H_{d-1/2}-\log 4\right)\cB_{IKJL}
\Big] \ .
\label{CC3}
\end{align}
In order to obtain a full set of the equations 
that relate the OPE data with the geometric data in the bulk, 
we have to work out the double OPE limit of 
the fout-point functions
\begin{align}
 \langle \cO_I(x_1)T_{ij}(x_2)\cO_K(x_3)T^{ij}(x_4)\rangle \ ,
~~~
\langle T_{ij}(x_1)T_{kl}(x_2)T^{ij}(x_3)T^{kl}(x_4)\rangle \ .
\nn
\end{align}
These are computed by promoting the bulk metric to a dynamical field,
which amounts to incorporating 
the bulk graviton exchange into (\ref{4ptf}).
We leave it for a future work.

(\ref{CC1}) shows that $C_{IK}^{(0)\,a}$
is regarded as an orthogonal matrix that relates the two
operator bases $IK$ and $a$. (\ref{CC2}) implies
that this diagonalizes the symmetric
matrix $\cA_{IKJL}-5\,\cB_{IKJL}$ with the eigenvalues proportional
to the anomalous dimensions $\gamma_{a}$.
$C_{IK}^{(1)\,a}$ is obtained by solving the linear equation (\ref{CC3}).


\section*{Acknowledgments}
We would like to thank Ken Kikuchi and Naoki Watamura 
for discussions.

\makeatletter
\renewcommand{\theequation}
{\Alph{section}.\arabic{equation}}
\@addtoreset{equation}{subsection}
\makeatother
\appendix

\section{Modified propagators}
\label{mGK}

Here, we make a brief review of how to derive the modified Green functions
used in this paper. 

The Green function of a bulk scalar field of mass $m$ in AdS$_{d+1}$
is defined as
\begin{align}
 (\Box_X-m^2)G_\Delta(X,Y)
=-\frac{1}{\sqrt{\gamma}}\,\delta^{d+1}(X-Y)
\ .
\label{def:green}
\end{align}
Fourier-tranforming $G_\Delta$ as
\begin{align}
 G_\Delta(X,Y)=\int \frac{d^dk}{(2\pi)^d}\,e^{i\vk\cdot(\vx-\vy)}\,
\wt G_\Delta(z,w,k) \ ,
\end{align}
$\wt G_\Delta$ is found to be solved as
\begin{align}
 \wt G_\Delta(z,w,k)=z^{d/2}\,\wt H_\Delta(z,w,k)
\end{align}
with $\wt H_\Delta$ defined by
\begin{align}
 \left(\cD_z-k^2 \right)\wt H_\Delta(z,w,k)
=-z^{d/2-1}\,\delta(z-w) \ .
\end{align}
Here, $\cD_z$ is the differential operator given by
\begin{align}
 \cD_z=\partial_z^2+\frac{1}{z}\partial_z-\frac{\nu^2}{z^2} \ ,
\end{align}
and $\nu=\sqrt{m^2+d^2/4}$\ . 
It is not diffucult to show that the complete set for $\cD_z$
is formed by the Bessel functions 
$\{J_\nu(\lambda z)\}_{\lambda\ge 0}$ with
the orthonormal condition given by
\begin{align}
\int_0^\infty dz\,z\,J_\nu(\lambda z)J_\nu(\lambda^\prime z)
=\frac{1}{\lambda}\delta(\lambda-\lambda^\prime) \ .
\end{align}
Then, $\wt H_\Delta(z,w,k)$ can be solved as a linear combination
of the Bessel function
\begin{align}
 \wt H_\Delta(z,w,k)=w^{d/2}\int_0^\infty d\lambda\,
\frac{\lambda}{\lambda^2+k^2}\,J_\nu(\lambda z) J_\nu(\lambda w) 
=\left\{
\begin{array}{c}
 w^{d/2}\, K_\nu(kz)I_\nu(kw) \ ,~~(z\ge w>0)\\
 w^{d/2}\, K_\nu(kw)I_\nu(kz) \ ,~~(w> z>0)\\
\end{array}
\right.
\ .
\end{align}
Here, $I_\nu$ is the modified Bessel function.

Now we define the modifed bulk-to-bulk propagator
\begin{align}
 G_\Delta^\epsilon(X,Y)\equiv G_\Delta(X,Y)
-(zw)^{d/2}\int \frac{d^dk}{(2\pi)^d}\,e^{i\vk\cdot(\vx-\vy)}\,
K_\nu(kz)K_\nu(kw)\,\frac{I_\nu(k\epsilon)}{K_\nu(k\epsilon)} \ .
\end{align}
This solves the equation (\ref{def:green}) because
\begin{align}
 (\cD_z-k^2)K_\nu(kz)=0 \ .
\end{align}
Then, it is easy to verify that $G_\Delta^\epsilon$ vanishes at the cut-off
surface $z=\epsilon$.

In order to define the modified bulk-to-boundary propagator
$K_\Delta^\epsilon(\vx,Y)$,
consider
\begin{align}
 \frac{\partial}{\partial w}G_\Delta^\epsilon(X,Y)
\bigg|_{z\ge\epsilon,w=\epsilon}
&=
(z\epsilon)^{d/2}\int\frac{d^dk}{(2\pi)^d}\,e^{i\vk\cdot(\vx-\vy)}\,
k\,\left[
K_\nu(kz)I^\prime(k\epsilon)
-
K_\nu(kz)K_\nu^\prime(k\epsilon)\,
\frac{I_\nu(k\epsilon)}{K_\nu(k\epsilon)}
\right]
\nn\\
&=
\epsilon^{d-1}
\int\frac{d^dk}{(2\pi)^d}\,e^{i\vk\cdot(\vx-\vy)}\,
\frac{(kz)^{d/2}\,K_\nu(kz)}{(k\epsilon)^{d/2}K_\nu(k\epsilon)} 
\nn\\
&\equiv 
\epsilon^{d-1}\cdot\epsilon^{\Delta-d}\,
K_\Delta^\epsilon(X,\vy) \ ,
\label{Kdelta}
\end{align}
with $\Delta=\nu+d/2$.
Here, we used the formula
\begin{align}
 I_\nu(z)K_\nu^\prime(z)-I_\nu^\prime(z)K_\nu(z)=-\frac{1}{z} \ .
\end{align}
It is found that
\begin{align}
 K_\Delta^\epsilon(X,\vy)\big|_{z=\epsilon}
=\epsilon^{d-\Delta}\,\delta^d(\vx-\vy) \ .
\label{Kdelta_eps}
\end{align}
As $\epsilon\to 0$, $K_\Delta^\epsilon(X,\vy)$ reduces  to
\begin{align}
 K_\Delta^\epsilon(X,\vy) \to
\frac{1}{\pi^{d/2}}\,\frac{\Gamma(\Delta)}{\Gamma(\Delta-d/2)}\,
\left(
\frac{z}{z^2+(\vx-\vy)^2}\right)^\Delta \ .
\label{K:eps0}
\end{align}
The following formulae are useful:
\begin{align}
\partial_zK_\Delta^\epsilon(X,\vy)\big|_{z=\epsilon}
&=
\epsilon^{d-\Delta}\int\frac{d^dk}{(2\pi)^d}\,
e^{i\vk\cdot(\vx-\vy)}\,
k\,
\partial_z\log(z^{d/2}K_\nu(z))\big|_{z=k\epsilon}\ ,
\\
 \int d^dy\, K_{\Delta}^\epsilon(X,\vy)&=\epsilon^{d-\Delta} \ .
\label{int:K}
\end{align}

%
%

%

%
\end{document}